# On Board Data Handling (OBDH) based on PC104

Haryono
Indonesian National Institute of Aeronautics and Space (LAPAN)
Jl. Cagak Satelit KM.04 Rancabungur – Bogor 16310
haryono@lapan.go.id

*Abstract—* Developing OBDH for satellite should consider the condition of the existing research facilities and resources in an institution. Many OBDH developments have been failed because not considering the capabilities of the institution. Considering to the capabilities of the institution is important, because it is major factor whether building OBDH can be realized successfully or not.

System Design OBDH that has great opportunities to success in our research environment is to concentrate on developing the software for the OBDH. The software must be supported with the appropriate hardware which has been recognized as space qualified. Therefore selection board which has space qualified is important method: it has been conducted in this research. To develop good software in the term of perspective programming, fast, standardize, testable, multitasking: Operating System has been implemented in this OBDH.

This research showed the OBDH development is pretty fast and more realizable to the limited institution resources. This research has produced an OBDH prototype in terms of hardware/board selection and software development.

**Keywords- On Board Data Handling, PC 104, Multitasking, Close loop Testing.**

## I. Introduction

In our institution the toughest obstacle in building the OBDH is about the hardware/board. It is because the human resources and facilities are limited, especially in terms of hardware manufacturing and hardware testing.

Many computation board developments for space have been conducted, most of them use very intensive resources and not easy to achieve in a limited resources (*Human and Facility Resources*), they use FPGA to achieve multitasking environments [1][2][3]. Simple OBDH development also have been conducted but difficult to fulfill the requirement when the operation of satellite is become complex [4]. It just uses microcontroller to handle many data from other sub systems, when many data need to be handled and complex operation need to be applied, simple OBDH is difficult to fulfill the requirement.

In order to achieve the goal to build an OBDH prototype which is faster, easy and can fulfill the requirement as much as possible: this research was conducted.

It is decided to use a hardware based on PC104 which is available in the market. PC104 was chosen, because it has a high flexibility and have clear standards, so it will be easier to develop the software. Hardware should fulfill the requirement; it should able to work in space. This research will concentrate on development the software of OBDH, development of hardware was still done but for minor part. The objective of this research is to develop the OBDH prototype in the limited resources with fast development and realizable to be implemented.

## II. Methods

The method to achieve the objectives is:

1. Understanding the hardware criteria for space and selecting of the board that is available in the market with space qualified board.

2. To develop the software of the OBDH: *Requirement Analysis* (List the satellite requirement to the OBDH). *System Design* (Design of the software which is possible to be implemented on the hardware, including the tool and the Operating system). *Implementation* (Building the software). *Integration* (Integrating the software and the hardware). And last is *Testing* .

## III. OBDH Hardware Criteria

Board selection method: PC104 was selected due to its availability in the market and its standard; many of them have been prepared for *Flight Model* (FM) at an affordable price below $1500 each board. The Board has conducted a test of temperature, vibration and radiation. Generally to make sure the selected OBDH board is space qualified; it should have been conducting a qualification testing. Qualification testing includes:

**1. Testing the thermal**: to find the resistance of OBDH to the thermal environment, it is cyclic environment like in orbit, the sunlight and eclipse phase.

**2. Testing vacuum**: to find the resistance of OBDH in a vacuum environment. Generally, a phenomenon that occurs in a vacuum environment is *Outgasing* of materials / components. *Outgasing* can cause the OBDH not working properly.

**3. Testing of vibration**: to determine whether OBDH still work even the vibration is occur, this is due to when launching the satellite will have more vibration.

**4. Testing Electromagnetic Interference (EMI):** to determine the level of electromagnetic emissions generated and its effect to satellite components.

Without doing all kinds of testing above, the hardware that is selected has already been testing, therefore the job of building OBDH will be minimized, the limited facility resources and human resources do not become an obstacle anymore. It is our method to fast in the development of OBDH, so that can minimize the time in development of the hardware.

In **Figure 1** showed the hardware that has been designed. It consists of several PC104 boards, one which is lowest rack as main board and others are extended board, below the detail of each board (some have been tested with XTEST (temperature test certificate) at -40C to + 85C operation:

1. MPL-MIP405-3-XTEST PC / 104-Plus using processor IBM405GPr PowerPC (up to 400MHz) with 2 x USB, there are 4x RS232 and 10/100 Ethernet on a header that is used as a debugging and programming, 128MB SDRAM and 8MB Flash soldered on board [5].
2. MPL-OSCI-8MIX-XTEST PC / 104-Plus is a module with 8 serial lines (RS232 / 422/485), which can be selected according to the required protocol [6].
3. MPL-IDE2CF-1A44i-XTEST is the board to put the Compact Flash that can be locked in order not to lose when subjected to large vibration [7].
4. PC 104 Custom, it has task to acquire the temperature and to measure the current and voltage of OBDH.

Selection to PC104 standard is because these standards have had a heritage that flies in the orbit. PC104 board is designed to facilitate the achievement of the objectives of computerization efficiently as needed.

PC104 is a board that has a standard in terms of Form and Bus [8]. With those kind of standard the development of the operating systems and integration is much easier. Bus communication using the standard (ISA and PCI) led to high compatibility with various operating systems available.

The main board to be used is MPL-MIP405-3. The board has been done vibration test, temperature test and radiation test [9]. The board has been widely used on a variety of satellite design [10] [11] and [12]. Extended Board include MPL-OSCI-8MIX all accompanied by a certificate temperature test. **Figure 1** is the integration of each board to the main board. Custom PC104 board is in the form of custom board will be custom designed to fit the needs of specific sub-systems.

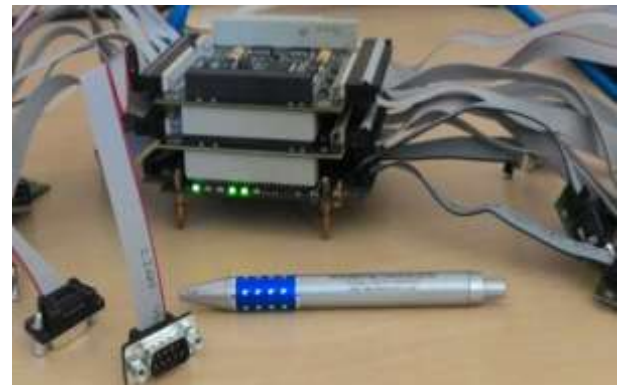

**Figure 1: Hardware of the OBDH**

Communication between boards is using PCI bus and IDE bus. The bus is a bus standard that is often used by the existing operating system (e.g. Linux / Windows). Voltage lines each board has been provided in the socket which has been provided by default via PC104 connector.

### IV. SOFTWARE DESIGN AND IMPLEMENTATION

Main board of OBDH is MPL-MIP405-3, it consist PPC 405 processor from IBM. To make the programming process is faster and easier, Linux Operating system was implemented. It just uses the core named *Linux Kernel*. The sub systems which have been connected are below:
1. Ground Segment (1 port).
2. Wheel Drive Electronics (WDE) (3 devices have been setup).
3. Star Sensor (2 devices have been set up).
4. Battery (1 Battery).
5. GPS (1 GPS).
6. Custom PC104 (1 port).

The task was divided based on the receiver data from other sub systems, the data flow is shown in **Figure 2**. Each Task is available each time the sub system sending the data to the OBDH. Each Task can process their task without any wonder to be interrupting other sub system. Ground Segment Communication Task has function to process the data or command from Ground Segment System. OBDH processes the data by extracting, formatting and to be determined what should be done based on which was received, either to be forwarded to other sub system or just being processed in the OBDH itself. WDE Communication Task, Star Sensor Communication Task, Battery Communication Task, GPS Communication Task and Custom PC104 Communication Task are act to process the data from other sub system. After processing is finish in each task, the task can decide whether to send the data to Ground Segment directly or keep in OBDH Memory.

Total port which has been used in OBDH is 9 ports, and can be expanded easily as required. All port should be available all time; therefore implementing operating system is increase the reliability. Operating system image generation from the beginning to the end has been talked in [13]. At here will focus on software development of the OBDH in operating system which has been prepared in PC104. Next discussion will talk about each software development implementation in OBDH.

**IV.1. Setup the Port**

Each sub system has different port to communicate with OBDH. Therefore it is necessary to setup the port according to the characteristic of Sub system. Below is the setup port to communicate between subs systems:

**IV.1. 1. Initiation of ports**

Port variable is where the port of OBDH that is used, for example: */dev/ttyS2, dev/ttyS3, /dev/ttyOS0, /dev/ttyOS2, /dev/ttyOS4, /dev/ttyOS6, /dev/ttyOS1*. *ttySx* is the port which are placed in Main OBDH. *ttyOSx* is the port which are placed in Extended OBDH.

```
int fd = open(port, O_RDWR | O_NOCTTY );
```

**IV.1. 2. Initiate the termios**

Before setting the speed and other setting need to prepare the *termios* in memory. *Termios* is struct that consist: *input mode flags, output mode flags, control mode flags, local mode flags, line discipline, control characters, input speed, output speed*. Below is the initiate of the termios in memory.

```
struct termios tty;
memset (&tty, 0, sizeof tty);
```

**IV.1. 3. Setting the Speed**

Speed is the baud rate, how much speed that is required to be able communication between sub systems. For example communication with Ground Segment uses 9600 bp

```
cfsetospeed (&tty, 9600);
cfsetispeed (&tty, 9600);
```

*cfsetospeed* is the output speed and *cfsetispeed* is the input speed.

**IV.1. 4. Setting the control characters**

To make easy in communication with other sub system, to read block each 1 character was setup. This is selected because easy to handle various data which are coming to OBDH. By using 1 character per one time received will able to calculate how many characters which are coming easily and also can easy to give a validation to character stream which are coming.

```
tty.c_cc[VMIN]  = 1;      // read block 1 char
tty.c_cc[VTIME] = 5;      // 0.5 seconds read
```

The OBDH has been received the data successfully without having an overlap, it is important to setting *VMIN/VTIME* properly, because it is having impact in capturing all the data in one burst without having an overlap. *"VMIN > 0 and VTIME > 0"*, this is the most common mode of operation, and consider VTIME to be an *intercharacter* timeout, not an overall one, this call should never return zero bytes, can be read in [14].

**IV.2. Create Tasks**

To create a task contains 2 parts. Below are parts in creating the Task:

**IV.2.1. Initiate the function**

The function should use a reference, because it will be passing to the *pthread_create*.

```
pthread_t tid[15];
void* TaskRxOsci1(void *arg);
```

**IV.2.2. Create the task**

*tid* is the Thread identifiers, each task will have an unique ID. After calling this code the *TaskRxOsci1* will have own thread and will be executed separately or in parallel manner.

```
pthread_create(&(tid[1]), NULL, & TaskRxOsci1,
                NULL);
```

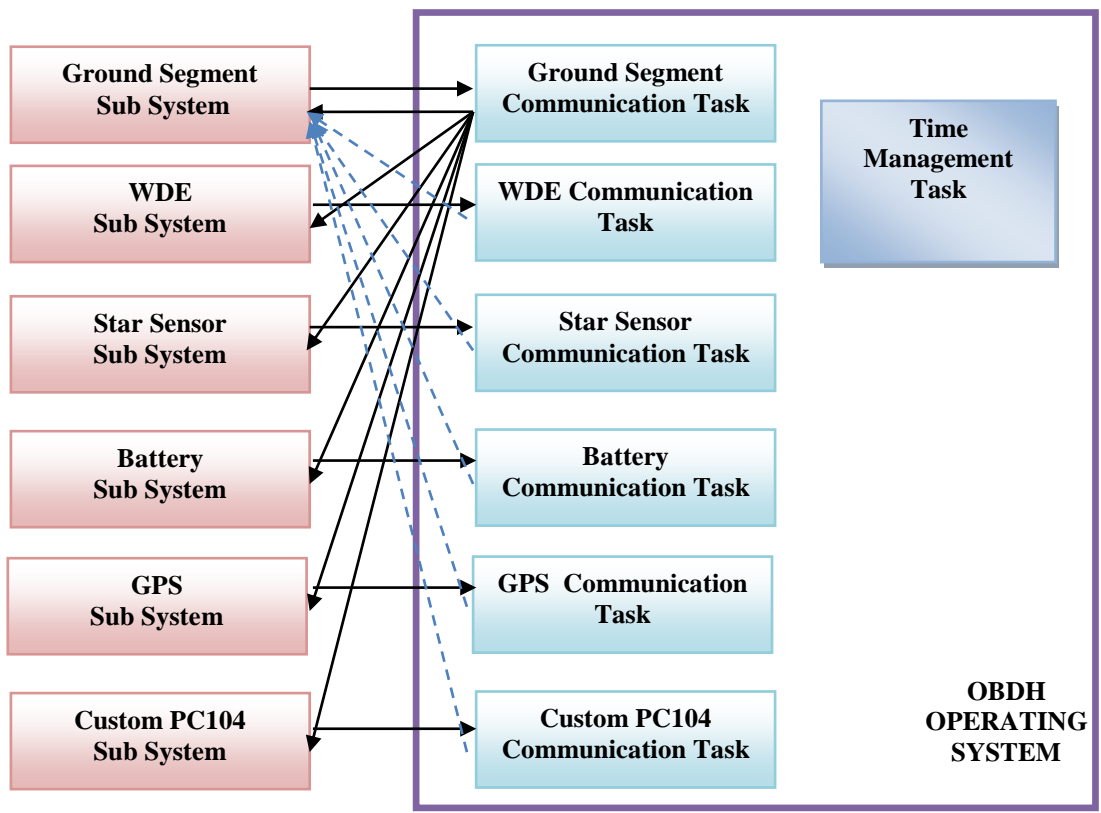

**Figure 2: Multitasking Data flow of OBDH**

### IV.3. Ground Segment (GS) Communication Task

In this task handle the communication between the ground segment and the OBDH. In **Figure 3** shows the communication data flow in GS communication task. Task will handle various data from GS transmit, it should able decoding the data and able to transmit to sub system destination correctly. At this task also format the data according to the sub system format.

In **Figure 4** shows about the snippets code for the Ground Segment (GS) Communication Task. It has *while* inside the function, because the function is executed parallel between other function and task, therefore it is not having impact to other task, because has independently time execution.

*read(PortRxMainBoard2,&rxChar,1)>0* will always in standby mode until **PortRxMainBoard2** receiving the data from GS. As **VMIN** is 1, therefore will read every one character and the character data will be hold by *rxChar* variable. **"#"** is the first character to be sent to the OBDH, so each time receives the **"#"** will reset the data. ''*rxMainBoard2Data[1] == 0x01*'' is checking to where sub system, the data will be sent, *0x01* is the ID of the sub system, it was connected to **PortRxMainBoard3**. So the rest of the data will be sent to the **PortRxMainBoard3** via

*write(PortRxMainBoard3,&rxMainBoard2Data[i],1)*.

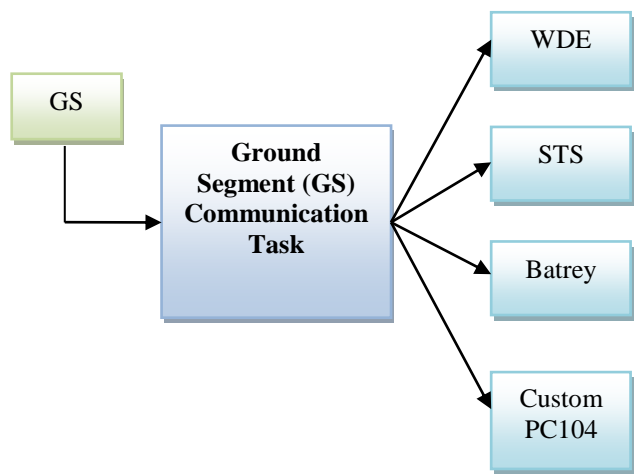

**Figure 3: Data Communication Flow in Ground Segment (GS) Communication Task**

```
void* TaskRxMainBoard2(void *arg)
{
while(1)
{
if (read(PortRxMainBoard2,&rxChar,1)>0)
{
    if(rxChar == '#')
    {
        index = 0;
        memset (&rxMainBoard2Data, 0, sizeof
            rxMainBoard2Data);
    }
    rxMainBoard2Data[index] = rxChar;
    index +=1;
    if(rxMainBoard2Data[0] == '#'
        && rxChar == '&')
    {
        if(rxMainBoard2Data[1] == 0x01)
        {
            for (i = 3 ; i < index -1 ; i++ )
            {
                write(PortRxMainBoard3,&
                    rxMainBoard2Data[i],1);
```

**Figure 4: Snippets Code for the Ground Segment (GS) Communication Task**

### IV.4. Wheel Drive Electronics (WDE) Task

**Figure 5** shows the Data Communication Flow in Wheel Drive Electronics Task. After WDE receiving the command data from the OBDH, WDE will replay what data is needed by the OBDH. WDE will know what data that is needed by the OBDH. OBDH will format the data properly to get the respond from WDE.

**Figure 6** shows the Snippets Code for Wheel Drive Electronics Task. *PortRxMainBoard3* is the port for WDE1, therefore the task read the data from *PortRxMainBoard3*. Each character is read then will be collected in *rxMainBoard3Data*. The last character in each stream data from WDE will be *0xAC*. Therefore after got the *0xAC* character, the data is ready to be used for the next process or operation, at here it just send to the Ground Segment (*PortRxMainBoard2*).

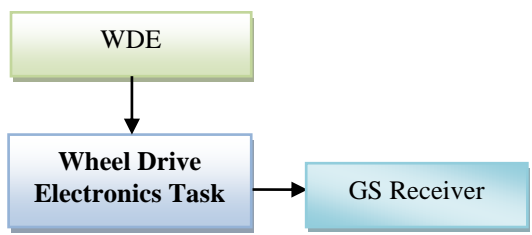

**Figure 5: Data Communication Flow in Wheel Drive Electronics Task**

```
void* TaskRxMainBoard3(void *arg)
{
while(1)
{
if (read(PortRxMainBoard3,&rxChar,1)>0)
{
    rxMainBoard3Data[index] = rxChar;
    index +=1;
    if(rxChar == 0xAC)
    {
        write(PortRxMainBoard2, "#",1);
        write(PortRxMainBoard2, &deviceId,1);
        write(PortRxMainBoard2,
            &rxMainBoard3Data,
                index);
        write(PortRxMainBoard2, "&",1);
```

**Figure 6: Snippets Code for Wheel Drive Electronics Task.**

### IV.5. Star Sensor Task

The task in star sensor is near same with the WDE task, the different in checking the data which are coming from Star Sensor. At **Figure 7** shows Snippets Code for Star Sensor Task, it will process the data after found the condition which is required. For example *if((FirstRxDataToSts1 == 0xA7 && IndexRxSTS1 == 3120)*, it will check whether the first data from Star sensor is *0xA7* and the number of stream data is *3120*, if correct will be consider the data and will be processed the data.

For the Battery, GPS, Custom PC104 Tasks will not so different to other task, the most different part will be in checking the format of the data each sub system which is connected to the OBDH. It should check the format of the data which is coming, if satisfied the condition which is required then will be processed the data for the operation used. After implemented in developing the software for the OBDH it can be known how essay to program the OBDH.

```
if((FirstRxDataToSts1 == 0x00 && IndexRxSTS1 ==
        152) ||
(FirstRxDataToSts1 == 0x01 && IndexRxSTS1 == 16) ||
(FirstRxDataToSts1 == 0xA0 && IndexRxSTS1 == 11) ||
    (FirstRxDataToSts1 == 0xA7 && IndexRxSTS1 ==
        3120) ||
(FirstRxDataToSts1 == 0xA8 && IndexRxSTS1 == 180)
        ||
(FirstRxDataToSts1 == 0x4D && IndexRxSTS1 == 8) ||
(FirstRxDataToSts1 == 0x02 && IndexRxSTS1 == 32) )
        {
```

**Figure 7: Snippets Code for Star Sensor Task**

## V. TESTING AND ANALYSIS

### V.1. Software Testing Analysis

Testing is conducted by the Functional Test and Qualification Test. Functional tests are unit tests and integration tests. Prior integration, unit tests has been conducted. Unit Test aims to ensure that each code works according to the desired. Any code that has been committed, it has been treated in the form of unit testing. Other testing that related to the software testing is testing data communications with other sub-systems. For example, communication with WDE, OBDH transmits a data to WDE, whether the data is able to be delivered properly to the WDE or not.

Unit testing is really important to make sure the code is working as expected and to be expected to maintain of previous working code. In this discussion will be talked about some example unit testing that has been done in Visual Studio. In **Figure 8** showed there is a code that aims to convert from Number to Binary string format [15]. In the unit testing, it is treating the *ConvertToBin* method code whether the method that has been writing up can work as expected. In **Figure 9** showed that *ConverToBin* method is called and gave some value in via parameter, in the first test number 8 has been given to *ConverToBin*, *ConverToBin* should return *"1000"*, other ways the testing is failed. It was happened to the next test, if send number *"3"* should return *"11"* and if send number *"2"* should return *"10"*. By doing unit test above it can be easily to maintain the existing code so that not break the existing code.

```
public class BinaryHelper
{
    public static string ConvertToBin(int x)
    {
        char[] bits = new char[32];
        int i = 0;

        while (x != 0)
        {
            bits[i++] = (x & 1) == 1 ? '1' : '0';
            x >>= 1;
        }

        Array.Reverse(bits, 0, i);
        return new string(bits);
    }
}
```

**Figure 8: ConvertToBin method: to convert the number to string in binary format.**

```
[TestClass()]
public class BinaryHelperTests
{
    [TestMethod()]
    public void ConvertToBinTest()
    {
        string stringBinary = BinaryHelper.ConvertToBin(8);
        Assert.AreEqual(stringBinary.Replace("\0", string.Empty), "1000");

        stringBinary = BinaryHelper.ConvertToBin(3);
        Assert.AreEqual(stringBinary.Replace("\0", string.Empty), "11");

        stringBinary = BinaryHelper.ConvertToBin(2);
        Assert.AreEqual(stringBinary.Replace("\0", string.Empty), "10");
    }
}
```

**Figure 9: ConvertToBinTest unit test: to test the convert ConvertToBin method**

### V.2. Close Loop Testing and Analysis

Close loop testing is to test the port connection to other sub system. The sub system which has been connected to the OBDH is listed in the **Table 1**. *PortRxMainBoardx* means the sub system is connected to the port which is resided in Main Board of OBDH. *PortRxOscix* means the sub system is connected to the port which is available in OSCI Board. Close loop testing is the test which has aim to test the connectivity each port in OBDH. The diagram of close loop is shown in the **Figure 10**. The data is coming from Computer via *rs232*, it will be received by *PortRxOsci3*, after that the data is processing in OBDH to be sent to the *PortRxOsci1* via *PortRxOsci0*, from *PortRxOsci1* will be transmitted to the *PortRxOsci2*, from *PortRxOsci2* will be transmitted to *PortRxOsci6*. From *PortRxOsci6* the data will be back to *PortRxOsci0* until reaching to the computer again. The data which was transmitted from computer should be exactly same when it was received by computer. This close loop test is done in a periodically time. More than 12 hours the test was conducted and got the correct result and passed.

**Table 1. Port connection mapping between OBDH and Sub system**

| Port In OBDH | Protocol | Sub System | Testing Method |
|---|---|---|---|
| PortRxMainBoard2 | RS232 | EGSE | Connected to EGSE |
| PortRxMainBoard3 | TTL | WDE 1 | Connected to WDE 1 |
| PortRxOsci0 | TTL | WDE 2 | Close loop hook 1st |
| PortRxOsci2 | TTL | WDE 3 | Close loop hook 2nd |
| PortRxOsci4 | RS422 | STS 1 | Connected to STS 1 |
| PortRxOsci6 | RS422 | STS 2 | Close loop hook 3rd |
| PortRxOsci1 | TTL | BATREY | Close loop hook 4th |
| PortRxOsci3 | RS232 | GPS | Close loop hook 5th |

### V.3. Integration Testing and Analysis

Integration Testing is testing the communication between OBDH and other sub system. In this testing will not connected to all sub systems, to avoid the complexity of harnessing, as in this development is just initial phase or just experiment. In **Table 2** showed the connection port in OBDH and the sub system connectivity. The data which has been received by OBDH should be correct. The scenario for this testing is by sending the command data from EGSE to request the data from WDE and STS. Beside that the test also gives a command to WDE to rotate the wheel. The test has been conducted and got the correct data and passed.

The tests which were conducted; the OBDH can go through each test successfully and getting passed result. Unit test is helping a programmer and software architect to maintain their code in the future, so that previous code will not be break because of new code is coming. The unit test is very important because the code will eventually goes many and many.

Close Loop Test was conducted and got success result. The OBDH can pass the close loop test, each data which is transmitted from computer was received by the computer exactly same. This test is helping us when the sub system is limited and of course this test will dismiss the complexity of cable wiring in a satellite. Integration testing, this test was integrating 3 sub systems; they are EGSE, WDE, and STS. Each of the sub system was able communicate to the OBDH successfully. This test will represent the ability of the OBDH to communicate to other sub system. From the various tests above the OBDH can be concluded that can work smoothly and successfully as per requirement.

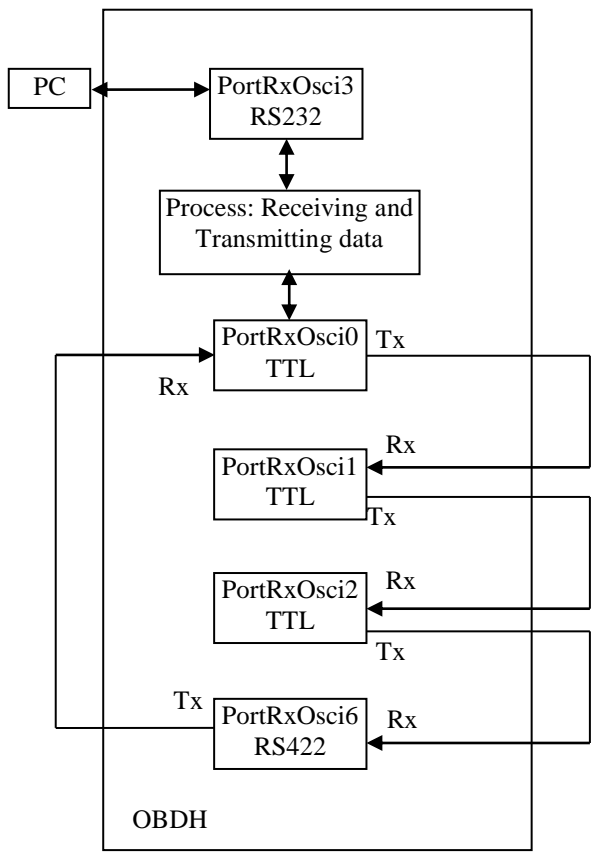

**Figure 10: Diagram of close loop test**

## VI. Conclusion and Future Work

The conclusion that can be obtained is by using PC104 and Operating system; the OBDH development is faster and more realizable to the limited resource of the institution. The OBDH prototype has been created within 1 year successfully, it emphasis in the software development only. The operating system image has been created and the software for the operations has also been created. Although it is not fully completed in a connection to other sub system because of limitation of the sub system device, it can represent how easy to develop the OBDH using PC104. The next research will integrate this OBDH with other sub systems in one table and to see the behavior and the operations.

## Acknowledgment

This project is supported by Satellite Centre - LAPAN, Mr Abdul Rahman as Head of Satellite Center LAPAN, Mr Abdul Karim as Head of Program Satellite LAPAN and my colleges Mr Fauzi and Mr Taufik. We would like to acknowledge for their support in this project.

## Author Biography

Haryono is researcher in Indonesian National Institute of Aeronautics and Space (LAPAN) - Satellite Technology Center. He is a Doctoral Degree in Computer Science from Gadjah Mada University, as a seasoned professional who has more than 10 years experience in Software Engineering.

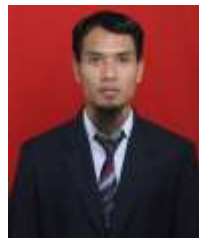